\documentstyle[preprint,pre,aps]{revtex}
\begin{document}
\draft
\title{Dynamic scaling behavior of an interacting monomer--dimer
model}
\author{Heungwon Park, Mann Ho Kim\cite{AAA} and Hyunggyu Park}
\address{Department of physics, Inha university, Inchon, 402-751,
Korea}
\date{\today}
\maketitle
\begin{abstract}
We study the dynamic scaling behavior of a monomer-dimer model with
repulsive interactions between the same species in one dimension.
With infinitely strong interactions the model exhibits a continuous
transition from a reactive phase to an inactive phase with two
equivalent absorbing states. This transition does not belong to the
conventional directed percolation universality class. The values of
dynamic scaling exponents are estimated by Monte Carlo simulations
for two distinct initial configurations, one near an absorbing state
and the other with an interface between two different
absorbing states. We confirm that the critical behavior is consistent with
that of the models with the mass conservation of modulo 2.
\end{abstract}


In recent years, nonequilibrium phase transitions occurring in
surface reaction models \cite{Liggett,Dickb,Harris,Grass79,ZGB,Ziff,%
Chopard,Bidaux,Aukrust,Dick89,Dick911,Jen90,Dick912,Jen91,Jen932,Jen933,%
Jen934,Albano92,Albano94,Park,Zhuo931,Zhuo932} have attracted great
interests, since they exhibit continuous transitions from a reactive
phase into an absorbing phase. Many investigations of models for
absorbing phase transitions have established a generally
accepted belief that continuous transitions into a single absorbing
state generically belong to one universality
class, the directed percolation (DP) class \cite{Jan,Grass82,Cardy}.
This
conjecture becomes extended to the multiple component models by the
argument of Grinstein {\em et al} \cite{Grin}.

Recently, a few models have appeared in the literature which are
known to be in a different universality class from DP.
Those are the model $A$ and $B$ of probabilistic cellular automa (PCA)
introduced by Grassberger {\it et al} \cite{Grass84,Grass89},
branching annihilating random walks with an even
number of offsprings (BAW) \cite{Taka,Sud,Jen931,Jen941},
nonequilibrium kinetic Ising models with two different
dynamics (NKI) \cite{Meny94,Meny95},
and an interacting monomer-dimer model (IMD) \cite{Park94}.
Critical behaviors of these models are different from DP but
seem to belong to the same universality class.
The PCA, BAW, and NKI models are basically single component models
(there are only two choices at a given site: vacant/occupied or
spin-up/spin-down),
while the IMD model is a multicomponent model (three choices:
vacant/monomer/dimer). The common feature of these models is that the
total number of particles (walkers in BAW, domain walls or kinks in
PCA, NKI, IMD) is conserved of modulo 2. But
there exist many types of kinks in the IMD model
due to the multi-component nature and each type
of kinks has no conservation law. So one needs to investigate
the IMD model in more detail.

In our previous study for the IMD model \cite{Park94},
the steady-state exponents are numerically determined as
$\beta=0.88(3)$, $\nu_\perp=1.83(3)$, $\nu_\parallel=3.17(5)$, which
agree well with the values of those exponents for the single component
models \cite{Grass89,Jen941,Meny94}.
In this study, we report the values of the dynamic
exponents for the IMD model and show that these values are
in excellent agreement with those of the BAW model with four
offsprings recently studied in detail by Jensen \cite{Jen941}.

The interacting monomer-dimer model (IMD) is a generalization of the
simple monomer-dimer model \cite{ZGB}
on a catalytic surface, in which particles of the same species
have nearest-neighbor repulsive interactions. This is parameterized
by specifying that a monomer ($A$) can adsorb at a nearest-neighbor
site of an already-adsorbed monomer (restricted vacancy) at a rate
$r_Ak_A$ with $0 \leq r_A \leq 1$, where $k_A$ is an adsorption rate
of a monomer at a free vacant site with no adjacent monomer-occupied
sites. Similarly, a dimer ($B_2$) can adsorb at a pair of restricted
vacancies ($B$ in nearest-neighbor sites) at a rate  $r_Bk_B$ with
$0 \leq r_B \leq 1$, where $k_B$ is an adsorption rate of a dimer at
a pair of free vacancies. There are no nearest-neighbor restrictions
in adsorbing particles of different species.
Here we will consider only the adsorption-limited reactions. A
nearest neighbor of the adsorbed $A$ and $B$ particles reacts
immediately, forms the $AB$ product, and desorbs the catalytic
surface.
Whenever there is an $A$
adsorption attempt at a vacant site inbetweeen an
adsorbed $A$ and an adsorbed $B$, we allow the $A$ to adsorb
and react immediately with the neighboring $B$, thus forming
an $AB$ product and  desorbing the surface \cite{inf}.
The case $r_A = r_B = 1$ corresponds to the ordinary noninteracting
monomer-dimer model which exhibits a first-order phase transition
between two saturated phases in one dimension. In the other limiting
case $r_A = r_B = 0$, the system has no fully saturated phases of
monomers or dimers, but instead two equivalent half-filled absorbing
states. These states comprise of only the monomers at the
odd- or even-numbered lattice sites. A dimer needs a pair of
adjacent vacancies to adsorb, so a state with
alternating sites occupied by monomers can be identified with an
absorbing state.

In this paper, we consider the one-dimensional IMD
with $r_A = r_B = 0$ for simplicity.
Then the system can be characterized by one parameter
$p = k_A/(k_A + k_B)$ of the monomer adsorption-attempt probability.
The dimer adsorption-attempt probability is given by $q = 1 - p $.
The order parameter of the system is the concentration of dimers
in the steady state, which vanishes algebraically as $p$ approaches
the critical probability $p_c$ from below.
Finite-size-scaling analysis of the static Monte Carlo data reveal
that this model behaves differently from the DP, but belongs to the same
universality class as the models in which the number of particles
(or kinks) are conserved of
modulo 2 \cite{Park94}.

We perform dynamic Monte Carlo simulations for the IMD model.
We start with two distinct initial configurations, one near an
absorbing state and the other with an interface between two
different absorbing states. Dynamic simulations with the former
initial configurations (conventional one) describe the evolution of
defect spreading on a nearly-absorbing space. We call this
``defect dynamics''. The other dynamics with the latter
initial configurations describe the evolution of interface spreading
between two different absorbing states. In contrast to the defect
dynamics the system can never enter an absorbing state. We
call this ``interface dynamics''.

In simulations for the defect dynamics, we start with a lattice occupied
by monomers at alternating sites except at the central vacant
site. Then the system evolves along the dynamic rules of the model.
After one adsorption attempt on the average per lattice site (one
Monte Carlo step), the time is incremented by one unit. A number of
independent runs, typically $2.5\times 10^5$,
are made up to 8000 time steps
for various values of $p$ near the critical probability $p_c$.
Most runs, however, stop earlier because the system gets into an
absorbing state. We measure the survival probability $P(t)$ (the
probability that the system is still active at time $t$),
the number of dimers $N(t)$ averaged over all
runs, and the mean-square distance of spreading $R^2 (t)$ averaged
over the surviving runs.
At criticality, the values of these quantities scale
algebraically in the long time limit \cite{Grass79}
\begin{eqnarray}
P(t) &\sim& t^{-\delta},\\
N(t) &\sim& t^{\eta},\\
R^2(t) &\sim& t^{z},
\end{eqnarray}
and double-logarithmic plots of these values against time
show straight lines. Off criticality, these plots show
some curvatures.
More precise estimates for the scaling exponents can be obtained
by examining the local slopes of the curves.
The effective
exponent $\delta(t)$ is defined as
\begin{equation}
-\delta(t) = \frac{\log \left[ P(t) / P(t/b) \right]}{\log ~b}
\end{equation}
and similarly for $\eta (t)$ and $z(t)$. In Fig.~1, we plot
the effective exponents against $1/t$ with $b =10$.
Off criticality these plots
show upward or downward curvatures. From Fig.~1, we estimate
$p_c  \simeq  0.5325(5)$ which is fully consistent with the result of
static Monte Carlo simulations \cite{error}.
The scaling exponent is given by the intercept
of the critical curve with the vertical axis. Our estimates for the
dynamic scaling exponents are
\begin{equation}
  \delta = 0.29(2),~~~ \eta = 0.00(2), ~~~ z = 1.34(20).
\end{equation}
These values agree well with those of the BAW with four
offsprings $(\delta=0.285(2), \eta=0.000(1), z=1.141(2))$ \cite{Jen941},
although the exponent $z$ shows very slow convergence.
In fact, the value of $z$ can be deduced from the values of
the steady-state exponents by
the scaling relation $z=2\nu_\perp$/ $\nu_\parallel$. Using
the previous steady-state results, we find $z\simeq 1.15$ which
is consistent with the above result within errors.
These values also satisfy the conventional hyperscaling relation,
$4\delta+2\eta=dz$ where $d$ is the spatial dimension \cite{Grass79}.

For the interface dynamics, we start with a pair of vacancies placed
at the central sites of a lattice and with monomers occupied at
alternating sites. In this case, the system never enter an absorbing
state, so that the survival probability exponent $\delta$
must be zero. 5000 independent runs are made during 8000 time steps
and we measure $N(t)$ and $R^2(t)$. In Fig.~2, we plot
the effective exponents $\eta(t)$ and $z(t)$ against $1/t$ with $b =10$.
The value of $p_c$ obtained from these plots is consistent with
the result from the defect dynamics.
Our estimates for the dynamic scaling exponents
are
\begin{equation}
  \eta = 0.285(20), ~~~~~~ z = 1.14(2).
\end{equation}
These values also agree well with those of the BAW with four
offsprings $(\delta=0.282(4), z=1.141(2))$ \cite{Jen941} and satisfy
the generalized hyperscaling relation, $2\delta+2\beta/\nu_\parallel
+2\eta=dz$ recently proposed by Mendes {\em et al} \cite{Mendes}.
Even though the
defect dynamics and the interface dynamics yield the different values
of dynamic exponents $\delta$ and $\eta$, their sum $\delta+\eta$
which is responsible for the growth of the number of kinks only in
surviving samples
seems to be the same. This property has been also observed in the
models with infinitely many absorbing states \cite{Mendes}.
In the IMD model,
the boundaries of the active region contact with one absorbing
state (defect dynamics) or two different absorbing states (interface
dynamics). But in the long time limit the active region becomes much
bigger in the surviving samples and local dynamics near boundaries
cannot distinguish which absorbing state is nearby. Therefore
the defect dynamics and the interface dynamics should give the same
result if only surviving samples are considered.

In summary, we have measured dynamic scaling exponents
of the interacting monomer-dimer model with infinitely strong
interactions by Monte Carlo simulations. This model exhibits
a continuous transition from a reactive phase to an inactive
phase with two equivalent absorbing states. Our results for
the dynamic exponents are fully consistent with those of the
branching annihilating walks with four offsprings which
conserve the number of paticles of modulo 2. Thus now we
firmly believe that these models belong to the same universality
class \cite{gamma}.

We wish to thank R.~Dickman and I.~Jensen for interesting
discussions. This work is supported in part by NON DIRECTED
RESEARCH FUND, Korea Research Foundation and by the BSRI, Ministry of
Education (Grant No.~95-2409).

\newpage
{\center\Large Figure Captions}
\begin{description}
   \item[Fig.\  1  :]  Plots of the effective exponents against
    $1/t$ for the defect dynamics. Five curves from top to bottom
    in each panel correspond to $p=0.5305,$ 0.5315, 0.5325, 0.5335,
    0.5345. Thick lines are critical lines ($p=0.5325$).
   \item[Fig.\  2  :]  Plots of the effective exponents against
    $1/t$ for the interface dynamics. Five curves from top to bottom
    in each panel correspond to $p=0.5285,$ 0.5305, 0.5325, 0.5345,
    0.5365. Thick lines are critical lines ($p=0.5325$).

\end{description}


\begin{references}
\bibitem{inf} If this process is not allowed, the IMD model
                possesses infinitely many absorbing states,
                which is currently under study.
\bibitem{error} We found a minor mistake in applying a local
                   dynamic rule of the IMD model in our previous
                   paper \cite{Park94}. Static Monte Carlo
                   simulations were reperformed with the correct
               dynamic rule. We found that the critical probability
               is shifted downward to 0.5325(5) but there was essentially
               no change in the values of the steady-state exponents.
\bibitem{gamma} We also checked the susceptibility exponent $\gamma$
             by static simulations. We found that $\gamma\simeq 0$ which
             is also consistent with the result for the BAW with four
             offsprings \cite{Jen941}.
\end{references}
\end{document}